%% file: main.tex
\documentclass[sigconf,natbib=true]{acmart}

\usepackage[utf8]{inputenc}
\usepackage{booktabs}
\usepackage{balance}
\usepackage{enumitem}
\usepackage{adjustbox}
\usepackage[most]{tcolorbox}

\usepackage{color}
\usepackage{rotating}
\usepackage{amsmath}
\usepackage{pifont}
\usepackage{array}
\usepackage{bbm}
\usepackage{multirow}
\usepackage{makecell}

\usepackage{colortbl} 
\usepackage[normalem]{ulem}
\useunder{\uline}{\ul}{}
\usepackage{graphicx}
\usepackage{subcaption}
\usepackage[linesnumbered,algoruled,boxed,lined]{algorithm2e}
\usepackage{siunitx}

\usepackage{booktabs}

\newcommand{\ie}{{\it i.e.}}
\newcommand{\eg}{{\it e.g.}}

\newcolumntype{H}{>{\columncolor{orange!30}\color{black}}c}

\copyrightyear{2025}
\acmYear{2025}
\setcopyright{cc}
\setcopyright{acmlicensed}
\acmConference[CIKM '25] {Proceedings of the 34th ACM International Conference on Information and Knowledge Management}{ November 10--14, 2025}{Seoul, Republic of Korea.}
\acmBooktitle{Proceedings of the 34th ACM International Conference on Information and Knowledge Management (CIKM '25), November 10--14, 2025, Seoul, Republic of Korea}
\acmISBN{979-8-4007-2040-6/2025/11}
\acmDOI{10.1145/3746252.3760952}

\begin{document}

\title{LLM-Enhanced Linear Autoencoders for Recommendation}
    
\author{Jaewan Moon} \authornote{Both authors contributed equally to this research.}
\affiliation{
  \institution{Sungkyunkwan University} \city{Suwon} \country{Republic of Korea}}
\email{jaewan7599@skku.edu}

\author{Seongmin Park} \authornotemark[1]
\affiliation{
  \institution{Sungkyunkwan University} \city{Suwon} \country{Republic of Korea}}
\email{psm1206@skku.edu}

\author{Jongwuk Lee}\authornote{Corresponding author}
\affiliation{
  \institution{Sungkyunkwan University} \city{Suwon} \country{Republic of Korea}}
\email{jongwuklee@skku.edu}



\input{sec-abstract}
\maketitle

\input{sec-introduction}
\input{sec-background}
\input{sec-method}

\input{sec-experiments}
\input{sec-conclusion}

\begin{acks}
    This work was partly supported by the Institute of Information \& communications Technology Planning \& evaluation (IITP) grant and the National Research Foundation of Korea (NRF) grant funded by the Korea government (MSIT) (No. IITP-RS-2019-II190421, IITP-RS-2022-II220680, IITP-2025-RS-2024-00360227, and NRF-RS-2025-00564083, each contributing 25\% to this research).
\end{acks}

\section*{GenAI Usage Disclosure}
We utilized NV-Embed-v2~\cite{Lee0XRSCP25NVEmbed} to encode the semantic item representation. GenAI tools were also used during the writing of the paper, but the authors are fully accountable for the content.

\balance
\input{reference.bbl}

\end{document}

%% file: sec-abstract.tex
\begin{abstract}\label{sec:abstract}

Large language models (LLMs) have been widely adopted to enrich the semantic representation of textual item information in recommender systems. However, existing linear autoencoders (LAEs) that incorporate textual information rely on sparse word co-occurrence patterns, limiting their ability to capture rich textual semantics. To address this, we propose \emph{\textbf{L$^3$AE}}, the first integration of LLMs into the LAE framework. L$^3$AE effectively integrates the heterogeneous knowledge of textual semantics and user-item interactions through a two-phase optimization strategy. (i) L$^3$AE first constructs a semantic item-to-item correlation matrix from LLM-derived item representations. (ii) It then learns an item-to-item weight matrix from collaborative signals while distilling semantic item correlations as regularization. Notably, each phase of L$^3$AE is optimized through closed-form solutions, ensuring global optimality and computational efficiency. Extensive experiments demonstrate that L$^3$AE consistently outperforms state-of-the-art LLM-enhanced models on three benchmark datasets, achieving gains of 27.6\% in Recall@20 and 39.3\% in NDCG@20. The source code is available at \url{https://github.com/jaewan7599/L3AE_CIKM2025}.

\end{abstract}

%% file: sec-introduction.tex
\section{Introduction}\label{sec:introduciton}

Recommender systems have evolved across various applications to address information overload. Their primary objective is to accurately predict a user’s preferences for unexperienced items based on users' past behavior. Collaborative filtering (CF) represents a prevalent approach that mines user–item interaction data to uncover latent collaborative signals for personalized recommendations~\cite{MoonJCCSL23, 0007YLPL23}. Large language models (LLMs) have recently emerged as powerful tools for deriving semantic representations from textual item attributes (\eg, titles, categories, brands, and descriptions) in recommender systems. Broadly, LLM-based approaches fall into two categories: (i) \emph{LLM-as-Recommender}~\cite{BaoZZWF023TALLRec, LiaoL0WYW024LLaRA, LCZZX23E4SRec}, which fine-tunes LLMs directly on recommendation tasks to serve as end-to-end models, and (ii) \emph{LLM-as-Extractor}~\cite{iclr/Sheng0ZCWC25AlphaRec, ZYYZLXMY25LLMInit, RenWXSCWY024RLMRec}, which leverages LLM-generated item representations as initial embeddings and fine-tunes conventional recommender models to capture collaborative interaction patterns.

\input{Figures_tex/Figure1_Pilot_Study}

Building on the LLM-as-Extractor paradigm, this paper focuses on linear autoencoders (LAEs). LAEs~\cite{MoonKL23RDLAE, Steck19EASE, Steck20edlae, HongCLKP24svdae, VancuraAKK22ELSA, park2025dan, park2025tale, ChoiKLSL21} learn an \emph{item-to-item weight matrix} $\mathbf{B} \in \mathbb{R}^{n \times n}$ by reconstructing the user-item interaction matrix $\mathbf{X} \in \{0,1\}^{m \times n}$ for $m$ users and $n$ items. While LAEs have demonstrated strong performance with minimal computational overhead, they rely solely on sparse interactions, resulting in suboptimal performance, particularly for long-tail items. Prior efforts~\cite{recsys/NingK12CSLIM, www/NingK12CSLIM, JeunenBG20CEASE} introduced auxiliary textual information by constructing a tag-item matrix $\mathbf{T} \in \{0, 1\}^{|\mathcal{V}| \times n}$ via multi-hot encoding over a vocabulary $\mathcal{V}$, and jointly reconstructing $\mathbf{X}$ and $\mathbf{T}$. However, these multi-hot encodings merely reflect the lexical co-occurrence of tags, failing to capture semantic similarities between textually distinct but conceptually similar items (\eg, \textit{`running shoes'} vs. \textit{`athletic sneakers'}).

To address this semantic gap, we investigate the first integration of LLMs into the LAE framework. We first construct a semantic-item matrix $\mathbf{F} \in \mathbb{R}^{d \times n}$ where each column represents an item's $d$-dimensional embedding vector obtained from LLM-derived textual attributes. Figure~\ref{fig:head_tail_performance} reveals that an LAE model with the semantic item matrix (\ie, LLM-EASE) outperforms both interaction-only (\ie, EASE) and multi-hot encoding models (\ie, CEASE and Add-EASE), with particularly pronounced gains on long-tail items. These results indicate that the existing study~\cite{JeunenBG20CEASE} overlooks the complementary nature of semantic and collaborative knowledge. While interaction data captures user preferences, semantic embeddings reveal crucial relationships between items.

In this paper, we propose a novel \emph{LLM-enhanced LAE (\textbf{L$^3$AE})} model that effectively integrates rich semantic representations with collaborative item signals in the LAE framework. Specifically, L$^3$AE operates in two phases to adequately consider the heterogeneous knowledge of both data sources. (i) It first encodes textual attributes into semantic embeddings $\mathbf{F}$ using LLMs and constructs a semantic-level item-to-item weight matrix that captures fine-grained item correlations; (ii) Inspired by knowledge distillation (KD)~\cite{HintonVD15KD, TianKI20RKD, AguilarLZYFG20RKD}, it then learns an item-to-item weight matrix from user-item interactions $\mathbf{X}$, regularized by the semantic correlation matrix to align collaborative learning with textual semantics. Notably, both phases are optimized through closed-form solutions, which guarantee global optimality and preserve computational efficiency. As shown in Figure~\ref{fig:head_tail_performance}, L$^3$AE achieves the highest overall performance while demonstrating superior tail item performance, effectively combining interaction and semantic knowledge. Experimental results demonstrate that L$^3$AE consistently outperforms existing LLM-integrated models across three benchmark datasets, achieving average improvements of 27.6\% in Recall@20 and 39.3\% in NDCG@20. The effectiveness of L$^3$AE is pronounced on long-tail items, bridging the semantic gap in sparse interaction settings.

Our key contributions are summarized as follows:

\begin{itemize}[leftmargin=5mm]
\item \textbf{Framework}: We formulate L$^3$AE, a novel LAE architecture that integrates LLM-derived semantic embeddings with CF, replacing conventional multi-hot encodings while retaining closed-form optimization.

\item \textbf{Model design}: We learn the item-to-item weight matrix from semantic knowledge of items using LLMs and unify semantic and collaborative signals via semantic-guided regularization.

\item \textbf{Evaluation}: Extensive experiments validate the superior performance of L$^3$AE on three datasets, with substantial gains on long-tail item recommendations.
\end{itemize}

%% file: Figures_tex/Figure1_Pilot_Study.tex
\begin{figure}[t!]
\centering
\begin{tabular}{c}
\includegraphics[width=0.47\textwidth]{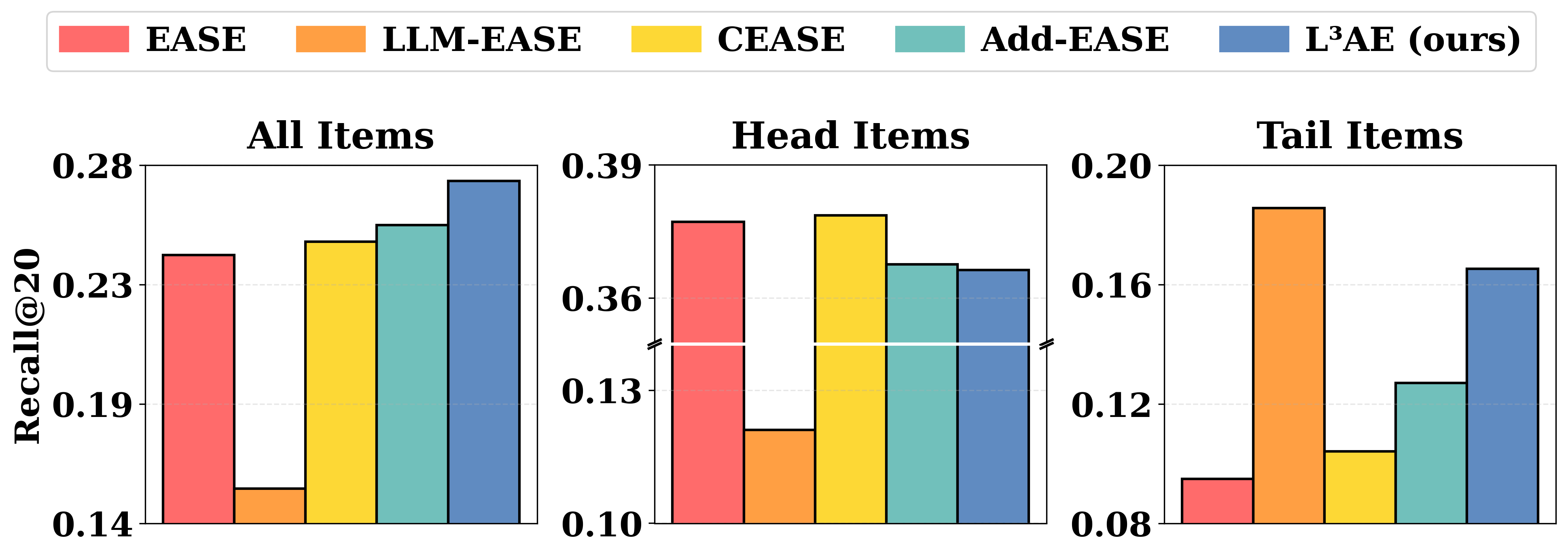} \\
\end{tabular}
\caption{Performance of head (top-20\% popular) and tail (the remaining 80\%) items on the Games dataset. LLM-EASE replaces the user-item interaction matrix with the semantic-item matrix from LLMs. Existing LAEs~\cite{JeunenBG20CEASE} are CEASE and Add-EASE, which utilize textual tag information.}
\label{fig:head_tail_performance}
\end{figure}

%% file: sec-background.tex
\section{Preliminaries}\label{sec:preliminaries}

\noindent
\textbf{Problem definition}. Assume that the user-item interaction is represented by a binary matrix $\mathbf{X} \in \{0, 1\}^{m \times n}$ for $m$ users and $n$ items. Here, $x_{ui}=1$ if user $u$ has interacted with item $i$, and $x_{ui}=0$ otherwise. The goal of recommender models is to identify the top-$k$ items that the user is most likely to prefer.

\vspace{0.5mm}
\noindent
\textbf{Linear autoencoders (LAEs)}. Given user-item interaction matrix $\mathbf{X} \in \{0,1\}^{m \times n}$, LAEs learn an \emph{item-to-item weight matrix} $\mathbf{B} \in \mathbb{R}^{n \times n}$ by reconstructing the interaction matrix $\mathbf{X}$. At inference, the prediction score $s_{ui}$ for user $u$ and item $i$ is computed as follows:
\begin{equation}
    s_{ui}=\mathbf{X}_{u*} \cdot \mathbf{B}_{*i},
\end{equation}
where $\mathbf{X}_{u*}$ and $\mathbf{B}_{*i}$ are the $u$-th row vector in $\mathbf{X}$ and the $i$-th column vector of $\mathbf{B}$, respectively.

As the simplest model, the objective function of EASE$^R$~\cite{Steck19EASE} is formulated by minimizing the reconstruction error with L$_2$ regularization similar to ridge regression~\cite{HoerlK00Ridge} and zero-diagonal constraints to remove self-similarity on the weight matrix $\mathbf{B}$:
\begin{equation}\label{eq:ease_objective}
    \min_{\mathbf{B}} \|\mathbf{X} - \mathbf{X} \mathbf{B}\|_F^2 + \lambda \|\mathbf{B}\|_F^2 \ \ \text{s.t.} \ \ \text{diag}(\mathbf{B})=0,
\end{equation}
where $\lambda$ controls the strength of L$_2$ regularization on $\mathbf{B}$.

Due to the convexity of the objective function, it yields the closed-form solution (See~\cite{Steck19EASE} for details.):
\begin{equation}\label{eq:ease_solution}
    \begin{split}
    \mathbf{B}_{EASE} &= \left(\mathbf{X}^{\top} \mathbf{X} + \lambda \mathbf{I}\right)^{-1}\left( \mathbf{X}^{\top}\mathbf{X} - \text{diagMat}(\boldsymbol{\mu}) \right) \\
                                &= \mathbf{I} - \mathbf{P} \cdot  \text{diagMat}( \textbf{1} \oslash \text{diag}(\mathbf{P}) ), \\
    \end{split}
\end{equation}
where $\mathbf{P} = (\mathbf{X}^{\top} \mathbf{X} + \lambda \mathbf{I})^{-1}$, $\textbf{1}$ and $\oslash$ are a vector of ones and the element-wise division operator, respectively. Lagrangian multipliers $\boldsymbol{\mu}$ enforce the zero-diagonal constraints, ensuring $\text{diag}(\mathbf{B})=0$.

\vspace{0.5mm}
\noindent
\textbf{Infusing textual information into LAEs}. Existing work~\cite{www/NingK12CSLIM, recsys/NingK12CSLIM, JeunenBG20CEASE} leverages auxiliary textual information of items by converting it into a multi-hot encoding format. Given a vocabulary $\mathcal{V}$ consisting of all tags (or words), a tag-item matrix $\mathbf{T} \in \{0, 1\}^{|\mathcal{V}| \times n}$ can be constructed analogously to the user-item interaction matrix $\mathbf{X}$. The existing study~\cite{JeunenBG20CEASE} proposed two methods to utilize both textual information and user-item interactions.

(i) \textbf{Collective method} employs a shared weight matrix $\mathbf{B}$ to reconstruct both the user-item interaction matrix $\mathbf{X}$ and tag-item matrix $\mathbf{T}$.
\begin{equation}\label{eq:collective_objective}
    \min_{\mathbf{B}} \|\mathbf{X} - \mathbf{X} \mathbf{B}\|_F^2 + \alpha \|\mathbf{T} - \mathbf{T} \mathbf{B}\|_F^2 + \lambda \|\mathbf{B}\|_F^2 \ \ \text{s.t.} \ \ \text{diag}(\mathbf{B})=0,
\end{equation}
where $\alpha$ controls the weight of the tag-item reconstruction term.

By stacking $\mathbf{X}$ and $\mathbf{T}$ into a matrix $\mathbf{X}' = \begin{bmatrix} \mathbf{X} \\ \sqrt{\alpha} \mathbf{T} \end{bmatrix}$, this objective function is reformulated to a similar form as Eq.~\eqref{eq:ease_objective} and yields the closed-form solution like Eq.~\eqref{eq:ease_solution}:
\begin{equation}\label{eq:stacked_collective_objective}
    \min_{\mathbf{B}} \|\mathbf{X'} - \mathbf{X'} \mathbf{B}\|_F^2 + \lambda \|\mathbf{B}\|_F^2 \ \ \text{s.t.} \ \ \text{diag}(\mathbf{B})=0.
\end{equation}
\begin{equation}\label{eq:stacked_collective_solution}
    \begin{split}
    \mathbf{B}_{Col} &= \mathbf{I} - \mathbf{P}_{Col} \cdot \text{diagMat}( \mathbf{1} \oslash \text{diag}(\mathbf{P}_{Col})),
    \end{split}
\end{equation}
where $\mathbf{P}_{Col} = (\mathbf{X'}^{\top} \mathbf{X'} + \lambda \mathbf{I})^{-1}$.

(ii) \textbf{Additive method} solves separate regression problems on the tag matrix $\mathbf{T}$ and the interaction matrix $\mathbf{X}$ to obtain two item-to-item weight matrices $\mathbf{C} \in \mathbb{R}^{n \times n}$ and $\mathbf{D} \in \mathbb{R}^{n \times n}$:
\begin{equation}\label{eq:additive_objective}
    \begin{split}
    \min_{\mathbf{C}} \|\mathbf{X} - \mathbf{X} \mathbf{C}\|_F^2 + \lambda_X \|\mathbf{C}\|_F^2 \ \ \text{s.t.} \ \ \text{diag}(\mathbf{C})=0, \\    
    \min_{\mathbf{D}} \|\mathbf{T} - \mathbf{T} \mathbf{D}\|_F^2 + \lambda_T \|\mathbf{D}\|_F^2 \ \ \text{s.t.} \ \ \text{diag}(\mathbf{D})=0,
    \end{split}
\end{equation}
where $\lambda_T$ and $\lambda_X$ adjust the strength of L$_2$ regularization for the interaction matrix and the tag matrix, respectively.

The solutions for $\mathbf{C}$ and $\mathbf{D}$ can be easily calculated by using Eq.~\eqref{eq:ease_solution}. Then, the final weight matrix $\mathbf{B}$ is formed by linear interpolation between two matrices $\mathbf{C}$ and $\mathbf{D}$:
\begin{equation}\label{eq:additive_solution}
    \mathbf{B}_{Add} = \beta \cdot \mathbf{C} + (1 - \beta) \cdot \mathbf{D},
\end{equation}
where $\beta$ controls weights for blending of two matrices $\mathbf{C}$ and $\mathbf{D}$.

The \textit{collective} method achieves global optimality through unified optimization, but it treats both heterogeneous data simultaneously within a single objective function. While the \textit{additive} method enables adaptive learning across heterogeneous data, but its na\"ive integration overlooks potential cross-source correlations.

%% file: sec-method.tex
\section{Proposed Method: L$^3$AE}\label{sec:methods}
We propose a \emph{\textbf{LL}M-enhanced \textbf{LAE} (\textbf{L$^3$AE})}. It consists of two phases: (i) constructing a semantic-level item-to-item matrix by leveraging semantics derived from LLMs and (ii) integrating heterogeneous knowledge via semantic-guided regularization.

\vspace{0.5mm}
\noindent
\textbf{Building semantic item representations using LLMs}.
While the multi-hot encoding strategy effectively captures lexical co-occurrences among tags, it inherently overlooks the underlying semantic similarities between them. This lexical-semantic gap limits the model's ability to leverage rich textual information. To bridge this gap, LLMs are employed to encode items into dense semantic representations. By projecting items into a semantic vector space, conceptually similar items are positioned closer together, enabling more effective modeling of semantic correlations. To encode the semantic item representation, we use a standard prompting method~\cite{Lee0XRSCP25NVEmbed, iclr/Sheng0ZCWC25AlphaRec}. The textual attributes are concatenated into a prompt without any explicit instructions: ``Title: <title>; Category: <category>; Brand: <brand>; Description: <description>''. This prompt is fed into LLMs, and the representation vector $f_i \in \mathbb{R}^{d \times 1}$ is obtained by averaging the final-layer token embeddings. By stacking these vectors for all items, we construct the semantic item matrix $\mathbf{F} \in \mathbb{R}^{d \times n}$.

\input{Figures_tex/Figure2_Singular_values}
\vspace{0.5mm}
\noindent
\textbf{Infusing heterogeneous knowledge into LAEs}.
A critical challenge remains: \textit{how can we effectively fuse the heterogeneous knowledge of user-item interactions and textual item semantics?} Although the \textit{collective} and the \textit{additive} methods in Eqs.~\eqref{eq:stacked_collective_solution} and~\eqref{eq:additive_solution} can utilize the semantic matrix $\mathbf{F}$ in place of the tag matrix $\mathbf{T}$, it remains unclear whether this simple replacement is appropriate. We conduct a pilot study to compare different characteristics between the interaction matrix $\mathbf{X}$ and the semantic matrix $\mathbf{F}$ through principal component analysis (PCA). Figure~\ref{fig:singular_values} shows the distributions of singular values for $\mathbf{X}$ and $\mathbf{F}$. The information of $\mathbf{F}$ is heavily concentrated in the top principal components, with the remaining dimensions near zero, indicating a low effective rank. In contrast, $\mathbf{X}$ exhibits a more gradual decay with sparsity-induced noise in its tail items~\cite{ShenWZSZLL21GFCF, PengLSM24sgfcf}.

Motivated by this observation, we propose a two-stage integration strategy that operates on item-item correlations rather than directly fusing raw data. This strategy enables each weight matrix to be regularized according to the distinct characteristics of its corresponding heterogeneous data source, while still deriving a globally optimal solution. In the first stage, we utilize $\mathbf{F}$ to construct a semantic correlation matrix $\mathbf{S}$ that captures the semantic structure among items. In the second stage, we estimate the final weight matrix $\mathbf{B}$ from interaction data, enhancing its objective with a semantic-guided regularization term that encourages $\mathbf{B}$ to align with $\mathbf{S}$. This design ensures that $\mathbf{B}$ effectively balances collaborative signals with rich semantic structure.

\vspace{0.5mm}
\noindent
\textbf{Construction of semantic item correlation (Phase 1)}. Instead of directly computing item similarity in the semantic space, we utilize the EASE framework~\cite{Steck19EASE}. Specifically, we learn a weight matrix $\mathbf{S}$ that captures the semantic correlation across items:
\begin{equation}\label{eq:semantic_ease}
    \min_{\mathbf{S}} \|\mathbf{F} - \mathbf{F} \mathbf{S}\|_F^2 + \lambda_F \|\mathbf{S}\|_F^2 \ \ \text{s.t.} \ \ \text{diag}(\mathbf{S})=0.
\end{equation}

Similar to~\eqref{eq:ease_solution}, it yields the closed-form solution:
\begin{equation}\label{eq:semantic_ease_solution}
    \begin{split}
    \mathbf{S} &= \left(\mathbf{F}^{\top} \mathbf{F} + \lambda_F \mathbf{I}\right)^{-1}\left( \mathbf{F}^{\top}\mathbf{F} - \text{diagMat}(\boldsymbol{\mu})\right) \\
     &= \mathbf{I} - \mathbf{P}_{F} \cdot  \text{diagMat}( \textbf{1} \oslash \text{diag}(\mathbf{P}_{F})),
    \end{split}
\end{equation}
where $\lambda_F$ adjusts the strength of L$_2$ regularization on $\mathbf{S}$, and $\mathbf{P}_{F} = (\mathbf{F}^{\top} \mathbf{F} + \lambda_F \mathbf{I})^{-1}$. Note that the weight matrix $\mathbf{S}$ leverages item semantic correlations rather than lexical matching.

\vspace{0.5mm}
\noindent
\textbf{Semantic-guided regularization (Phase 2)}. Inspired by knowledge distillation (KD)~\cite{HintonVD15KD, TianKI20RKD, AguilarLZYFG20RKD}, we learn the item-to-item weight matrix $\mathbf{B}$ via semantic-guided regularization using the pre-computed semantic matrix $\mathbf{S}$. L$^3$AE allows each source to receive its optimal L$_2$ regularization weight, adjusting the degree of regularization.

We formulate the objective function for learning $\mathbf{B}$ by extending Eq.~\eqref{eq:ease_objective} with a distillation term $\|\mathbf{B} - \mathbf{S}\|_F^2$, which minimizes the discrepancy between $\mathbf{B}$ and $\mathbf{S}$ in Eq.~\eqref{eq:semantic_ease_solution}:
\begin{equation}\label{eq:distill_objective}
    \min_{\mathbf{B}} \|\mathbf{X} - \mathbf{X} \mathbf{B}\|_F^2 + \lambda_X \|\mathbf{B}\|_F^2 + \lambda_{KD} \|\mathbf{B} - \mathbf{S}\|_F^2 \ \ \text{s.t.} \ \ \text{diag}(\mathbf{B})=0,
\end{equation}
where $\lambda_X$ controls the strength of L$_2$ regularization on $\mathbf{B}$ and $\lambda_{KD}$ governs the strength of the distillation term. This formulation encourages $\mathbf{B}$ to simultaneously capture collaborative signals from $\mathbf{X}$ and semantic relationships among items distilled from $\mathbf{S}$. When $\lambda_{KD}=0$, Eq.~\eqref{eq:distill_objective} simplifies to Eq.~\eqref{eq:ease_objective}, yielding LAEs that rely solely on interaction data (\ie, EASE$^R$~\cite{Steck19EASE}).

Solving the constrained optimization problem in Eq.~\eqref{eq:distill_objective} yields the following closed-form solution:
\begin{align}\label{eq:distill_solution}
    \mathbf{B}_{L^3AE}
    &= \left(\mathbf{X}^{\top} \mathbf{X} + (\lambda_{KD} + \lambda_X) \mathbf{I}\right)^{-1}
    \left( \mathbf{X}^{\top}\mathbf{X} + \lambda_{KD} \mathbf{S} - \text{diagMat}(\boldsymbol{\mu}) \right) \nonumber \\
    &= \mathbf{I} + \lambda_{KD} \mathbf{P}_{KD} \cdot \mathbf{S} - \mathbf{P}_{KD} \cdot  \text{diagMat}(\boldsymbol{\mu}), 
\end{align}
where $\mathbf{P}_{KD}=\left(\mathbf{X}^{\top} \mathbf{X} + (\lambda_{KD} + \lambda_X) \mathbf{I}\right)^{-1}$, $\boldsymbol{\mu} = \text{diag}(\textbf{1} + \lambda_{KD} \mathbf{P}_{KD} \cdot \mathbf{S}) \oslash \text{diag}(\mathbf{P}_{KD})$. In Eq.~\eqref{eq:distill_solution}, $\lambda_{KD}$ not only controls the influence of semantic correlations but also contributes to the regularization for interaction data (\ie, an equivalent role to $\lambda$ in Eq.~\eqref{eq:ease_solution}). Consequently, the regularization strength for interaction data $\mathbf{X}$ becomes $\lambda_{KD}+\lambda_X$.

%% file: Figures_tex/Figure2_Singular_values.tex
\begin{figure}[t!]
\centering
\begin{tabular}{cc}
\includegraphics[width=0.22\textwidth]{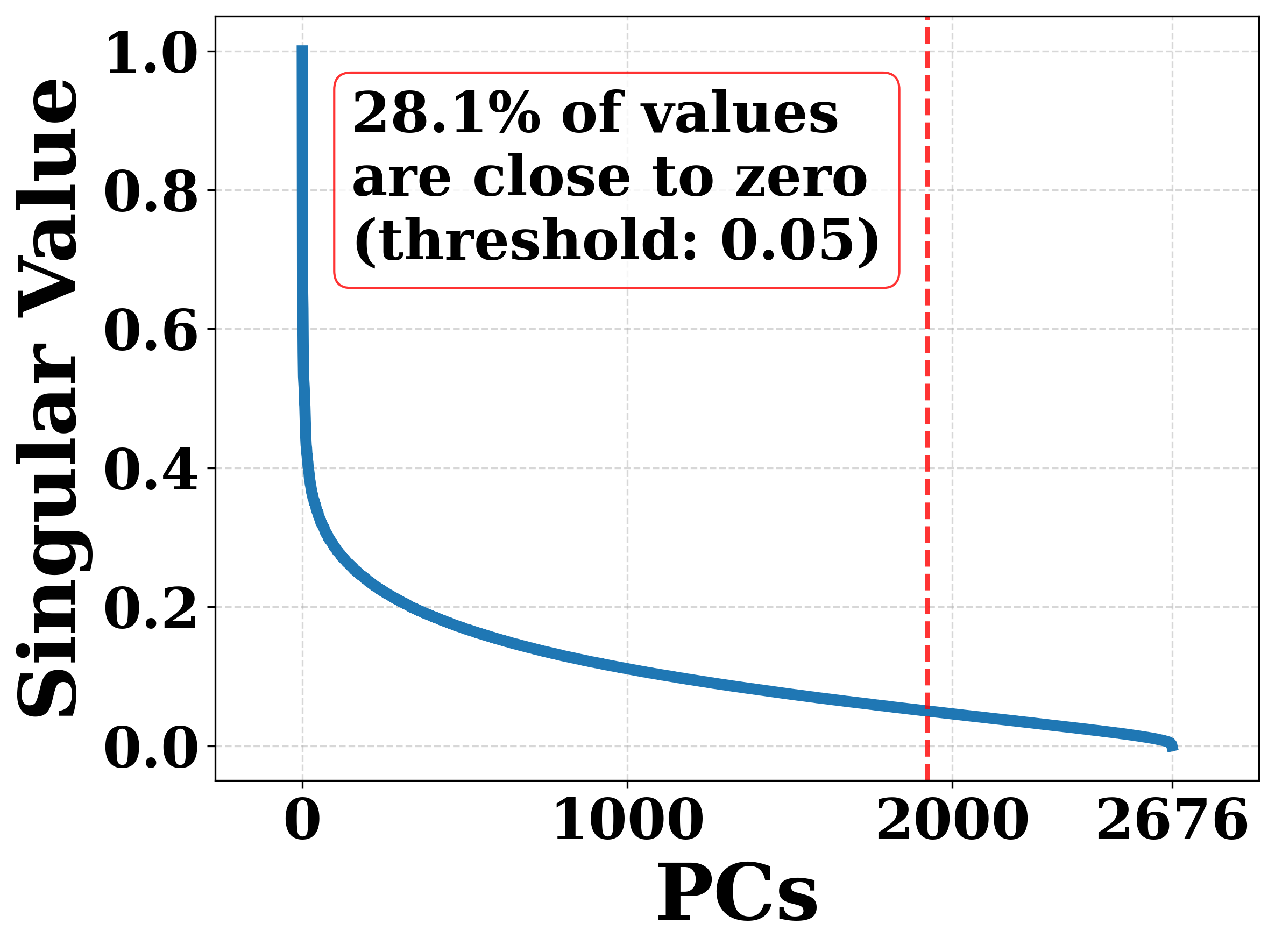} &
\includegraphics[width=0.22\textwidth]{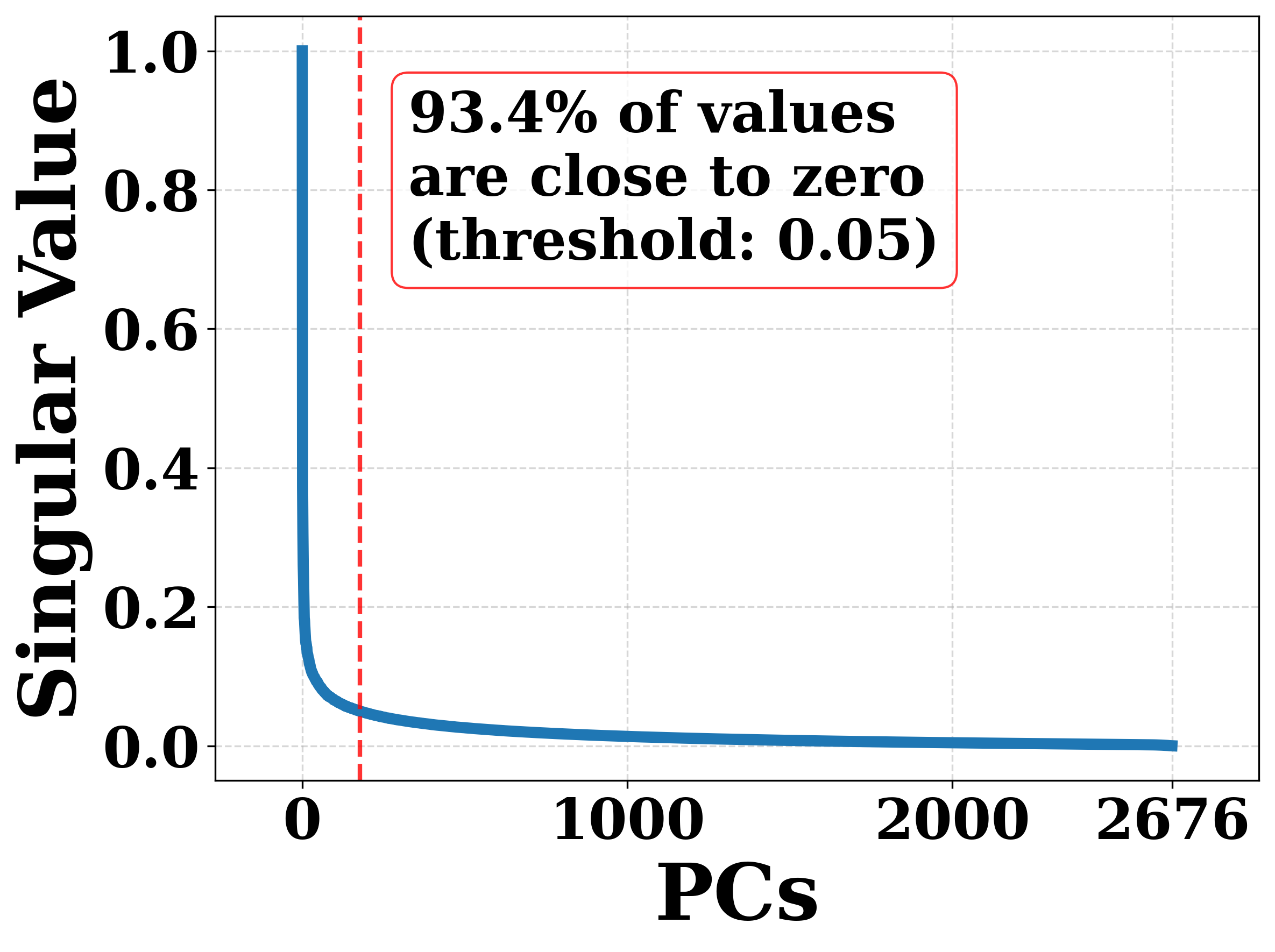} \\
(a) Interaction matrix $\mathbf{X}$ & (b) Semantic matrix $\mathbf{F}$ \\
\end{tabular}
\caption{Normalized singular values of interaction matrix $\mathbf{X}$ and semantic matrix $\mathbf{F}$ on Games, where the number of items is 2,676. We also observe similar trends on other datasets. }
\label{fig:singular_values}
\end{figure}

%% file: sec-experiments.tex
\section{Experiments}\label{sec:experiments_setup}

\input{Tables/Tab1_Data_Statistics}
\input{Tables/Tab2_Overall_Performance}

\subsection{Experimental Setup}
\subsubsection{Datasets}
We employ three Amazon 2023 datasets\footnote{\url{https://amazon-reviews-2023.github.io/}}~\cite{hou2024bridging}: Games (5.2K users, 2.7K items, 86K interactions, and sparsity: 99.39\%), Toys (14.8K users, 13.4K items, 251K interactions, and sparsity: 99.87\%) and Books (25.3K users, 31.0K items, 641K interactions, and sparsity: 99.92\%). Following existing work~\cite{YuY00CN22SimGCL, WuWF0CLX21sgl}, we retain interactions with ratings above 3, apply 10-core filtering, and split each dataset into training, validation, and test sets in an 8:1:1 ratio. The statistics of datasets are summarized in Table~\ref{tab:statistics}.

\subsubsection{Evaluation protocols} We employ the \textit{average-over-all} evaluation across all items a user has not interacted with to accurately measure each model's performance. We report two widely used metrics: Recall@$k$ (R@$k$) and NDCG@$k$ (N@$k$) with $k=\{10, 20\}$. R@$k$ quantifies the fraction of relevant items retrieved, and N@$k$ accounts for both the relevance and ranking position of the preferred items within the top-$k$ recommendation list.

\subsubsection{Competing models}. 
We compare our method against five non-linear models (\ie, LightGCN~\cite{0001DWLZ020LightGCN}, SimGCL~\cite{YuY00CN22SimGCL}, RLMRec-Con~\cite{RenWXSCWY024RLMRec}, RLMRec-Gen~\cite{RenWXSCWY024RLMRec}, and AlphaRec~\cite{iclr/Sheng0ZCWC25AlphaRec}) and seven linear models (\ie, cosine similarity as the item-to-item similarity matrix, EASE~\cite{Steck19EASE}, GF-CF~\cite{ShenWZSZLL21GFCF}, BSPM~\cite{ChoiHPC23BSPM}, SGFCF~\cite{PengLSM24sgfcf}, CEASE~\cite{JeunenBG20CEASE}, and Add-EASE~\cite{JeunenBG20CEASE}). Each model category is classified based on the training features it leverages: \textit{interaction} features derived from the user-item interaction matrix (\ie, LightGCN, SimGCL, EASE, GF-CF, BSPM, and SGFCF),  \textit{multi-hot} encoding (\ie, CEASE and Add-EASE), and \textit{semantics} representing LLM-derived information (\ie, RLMRec-Con, RLMRec-Gen, AlphaRec, Cos., EASE and L$^3$AE).

\subsubsection{Implementation details}.
We conduct all experiments with NVIDIA A6000 and Intel Xeon Gold 6226. Since L$^3$AE is agnostic to LLM architecture, we adopt NV-Embed-v2\footnote{https://huggingface.co/nvidia/NV-Embed-v2}~\cite{Lee0XRSCP25NVEmbed}, LLaMA-3.2-3B\footnote{https://huggingface.co/meta-llama/Llama-3.2-3B}~\cite{abs-2407-21783LLaMA3}, and Qwen3-Embedding-8B\footnote{https://huggingface.co/Qwen/Qwen3-Embedding-8B}~\cite{abs-2506-05176Qwen3Embedding}. Following AlphaRec~\cite{RenWXSCWY024RLMRec}, we obtain LLM-derived user embeddings for existing LLM-enhanced methods by averaging the embeddings of each user's interacted items from the training set.

For non-linear models, we use the Adam optimizer with a learning rate of 0.001, a batch size of 4096, and a hidden dimension of 32, applying early stopping with a patience of 50 based on the validation R@20. All results of non-linear models are averaged over five runs, and significance tests are conducted between L$^3$AE and non-linear models across these five runs. For RLMRec~\cite{RenWXSCWY024RLMRec}, we adopt SimGCL~\cite{YuY00CN22SimGCL} as the backbone. We determine the hyperparameters for each model through a grid search following the authors' guidelines.

For LAEs~\cite{Steck19EASE, JeunenBG20CEASE} including L$^3$AE, we search $\lambda$, $\lambda_X$ and $\lambda_F \in \{0.1, 0.5, 1, 5, ..., 1000\}$, $\lambda_{KD} \in \{10, 20, ..., 100, 150, ..., 300\}$. For the collective method, we search $\alpha \in \{0.1, 0.5, 1, 2, 3, 4, 5\}$, and for the addictive method, we search $\beta \in \{0.2, 0.4, 0.6, 0.8\}$. To prevent over-regularization of interaction data, we first determine the optimal regularization weight $\lambda$ for interaction data using Eq.~\eqref{eq:ease_solution}, then enforce the constraint $\lambda=\lambda_{KD}+\lambda_X$ for L$^3$AE to maintain appropriate regularization strength across both data sources.

\subsection{Experimental Results}\label{sec:results}

\noindent
\subsubsection{Overall performance}
Table~\ref{tab:overall_performance} reports performance on three real-world datasets with the NV-Embed-v2~\cite{Lee0XRSCP25NVEmbed} backbone models. We highlight four key findings:

\noindent
\textbf{(i)}
L$^3$AE consistently achieves the highest performance across all datasets. Specifically, L$^3$AE outperforms AlphaRec~\cite{iclr/Sheng0ZCWC25AlphaRec}, achieving average gains of 29.1\% and 39.8\% in R@20 and N@20, respectively, while surpassing EASE~\cite{Steck19EASE} by 14.7\% and 15.3\% in the same metrics. Moreover, L$^3$AE shows substantial gains over multi-hot encoding (\ie, CEASE and Add-EASE~\cite{JeunenBG20CEASE}), demonstrating that LLM representations contain semantically rich signals beneficial for CF.

\noindent
\textbf{(ii)} LLM-enhanced methods (\eg, AlphaRec and L$^3$AE) outperform interaction-only methods (\eg, SimGCL~\cite{YuY00CN22SimGCL} and SGFCF~\cite{PengLSM24sgfcf}). Among non-linear methods, AlphaRec demonstrates superior performance.

\noindent
\textbf{(iii)} Linear models consistently outperform non-linear models, with performance gaps widening as data sparsity increases (Games $\rightarrow$ Toys $\rightarrow$ Books). Compared to AlphaRec, L$^3$AE achieves performance gains of 10.3\%, 33.3\%, and 43.7\% in R@20 on the Games, Toys, and Books datasets, respectively. This corroborates that linear models generalize better in sparse environments due to their structural simplicity and resistance to overfitting.

\noindent
\textbf{(iv)} When relying solely on LLM-derived semantics, EASE surpasses the cosine similarity of the representation vectors. Thus, our semantic-guided regularization leverages the weight matrix of EASE rather than relying on the cosine similarity of the representations.

\subsubsection{Performance over fusion methods}
Table~\ref{tab:fusion_performance} reports performance comparison across three fusion methods: LLM-CEASE, LLM-Add-EASE, and L$^3$AE. The tag-item matrix $\mathbf{T}$ of CEASE or Add-EASE is replaced with the semantic-item matrix $\mathbf{F}$ of L$^3$AE. L$^3$AE consistently outperforms the other fusion variants, with average gains of 1.6\% in both N@20 and R@20 across all datasets, and up to 4.5\% and 3.4\% gains over Add-EASE on Toys. This confirms that our fusion scheme effectively infuses heterogeneous knowledge into LAEs.

\input{Tables/Tab3_Fusion_Performance}
\input{Figures_tex/Figure3_Hyperparameter_Sensitivity}
\input{Figures_tex/Figure4_Performance_over_Various_LLMs}

\subsubsection{Hyperparameter sensitivity}
Figure~\ref{fig:hyper_sensitivity} shows performance of L$^3$AE over varying regularization weights $\lambda_{KD}$, $\lambda_F$, and $\lambda_X$. We analyze $\lambda_{KD}$ while maintaining the constraint $\lambda=\lambda_{KD}+\lambda_X$ to isolate the effect of the semantic-guided regularization, where $\lambda$ is the ideal regularization weight for interaction data using EASE. In contrast, we relax this constraint to $\lambda_X$ to examine the effect of over-regularization on the interaction component. We observe that each weight shows a distinct optimal value. Interestingly, performance degrades sharply as $\lambda_{KD} + \lambda_X$ deviates from $\lambda$ (Figure~\ref{fig:hyper_sensitivity}(b)), validating our regularization strategy for interaction data.

\subsubsection{Performance on diverse LLM backbone models}
Figure~\ref{fig:performance_over_various_LLMs} compares three models (\ie, cosine similarity, semantic-only EASE, and L$^3$AE) built on three LLM backbones with different parameter sizes (\ie, LLaMA-3.2-3B~\cite{abs-2407-21783LLaMA3}, NV-Embed-v2-7B~\cite{Lee0XRSCP25NVEmbed}, and Qwen3-Embedding-8B) on Games, Toys, and Books. For all backbones, we obtain semantic item representations following the procedure in Section~\ref{sec:methods}. Detailed performance with LLaMA-3.2-3B and Qwen3-Embedding-8B are reported in Tables~\ref{tab:overall_llama} and~\ref{tab:overall_qwen}, respectively.

We observe merely a weak correlation between the number of parameters of LLMs and performance. Qwen3-Embedding-8B underperforms the smaller NV-Embed-v2-7B and is comparable to or even slightly worse than LLaMA-3.2-3B. This suggests that pretraining data and domain alignment matter more than model scale. Notably, NV-Embed-v2’s pretraining set includes e-commerce corpora such as AmazonReviews~\cite{McAuleyL13AmazonReviews} and AmazonCounterfactual~\cite{abs-2104-06893AmazonCounterfactual}, which appears to yield more informative item semantic representations. Across both datasets, cosine similarity with Qwen3-Embedding-8B exceeds that with LLaMA-3.2-3B. However, L$^3$AE achieves slightly higher performance with LLaMA-3.2-3B than with Qwen3-Embedding-8B, and semantic-only EASE likewise favors LLaMA-3.2-3B. This implies that, compared with simple covariance proximity score (\ie, cosine similarity), EASE's precision (\ie, inverse-covariance) score better captures the semantic space’s downstream suitability from a graphical model perspective~\cite{Steck19EASE}.

\input{Tables/Tab4_Overall_Performance_LLaMA}
\input{Tables/Tab5_Overall_Performance_Qwen3}

%% file: Tables/Tab1_Data_Statistics.tex
\begin{table}[t]
\centering
\caption{Dataset statistics of three Amazon review datasets.}
\label{tab:statistics}
\begin{tabular}{l|cccc}
\toprule
Dataset     & \# Users  & \# Items  & \# Ratings    & Density        \\
\midrule
Games       & 5,222     & 2,676     & 85,690        & \num{6.2e-3}    \\
Toys        & 14,750    & 13,358    & 250,509       & \num{1.3e-3}    \\
Books       & 25,300    & 30,966    & 640,901       & \num{8.2e-4}    \\
\bottomrule
\end{tabular}
\end{table}

%% file: Tables/Tab2_Overall_Performance.tex
\begin{table*}[t]
\caption{Performance comparison across three datasets with the NV-Embed-V2 backbone model. Bold indicates the best performance within each model category. * denotes statistically significant gains of L$^3$AE over the best non-linear model ($p < 0.0001$ for two-tailed t-test). }
\centering
\begin{adjustbox}{width=1\textwidth}
\label{tab:overall_performance}
\begin{tabular}{cccccc|cccc|cccc}
\toprule
\multicolumn{1}{c|}{\multirow{2}{*}{{\makecell[c]{Training \\ Features}}}} & \multicolumn{1}{c|}{\multirow{2}{*}{{Model}}}
& \multicolumn{4}{c|}{{Games}} & \multicolumn{4}{c|}{{Toys}} & \multicolumn{4}{c}{{Books}} \\
\multicolumn{1}{c|}{} & \multicolumn{1}{c|}{} & R@10 & R@20 & N@10 & N@20   & R@10 & R@20 & N@10 & N@20     & R@10 & R@20 & N@10 & N@20 \\
\midrule
\multicolumn{14}{c}{\textit{{Non-linear recommendation models}}} \\
\midrule
\multicolumn{1}{c|}{\multirow[c]{2}{*}{Interaction}}
& \multicolumn{1}{c|}{LightGCN}                         & 0.1453 & 0.2199 & 0.0799 & 0.0997     & 0.0520 & 0.0811 & 0.0281 & 0.0359     & 0.0973 & 0.1456 & 0.0566 & 0.0701 \\
\multicolumn{1}{c|}{} & \multicolumn{1}{c|}{SimGCL}     & 0.1510 & 0.2286 & 0.0831 & 0.1037     & 0.0611 & 0.0914 & 0.0338 & 0.0419     & 0.1122 & 0.1631 & 0.0661 & 0.0803 \\
\midrule
\multicolumn{1}{c|}{\multirow[c]{3}{*}{\makecell[c]{Interaction\\+ Semantics}}}
& \multicolumn{1}{c|}{RLMRec-Con}                       & 0.1635 & 0.2431 & 0.0908 & 0.1119     & 0.0714 & 0.1088 & 0.0392 & 0.0491     & 0.1157 & 0.1668 & 0.0686 & 0.0830 \\
\multicolumn{1}{c|}{} & \multicolumn{1}{c|}{RLMRec-Gen} & 0.1607 & 0.2437 & 0.0890 & 0.1111     & 0.0713 & 0.1079 & 0.0391 & 0.0488     & 0.1184 & \textbf{0.1728} & 0.0697 & \textbf{0.0849} \\
\multicolumn{1}{c|}{} & \multicolumn{1}{c|}{AlphaRec}   & \textbf{0.1677} & \textbf{0.2482} & \textbf{0.0961} & \textbf{0.1175}     & \textbf{0.0794} & \textbf{0.1180} & \textbf{0.0440} & \textbf{0.0542}     & \textbf{0.1194} & {0.1676} & \textbf{0.0705} & {0.0841} \\
\midrule
\multicolumn{14}{c}{\textit{{Linear recommendation models}}} \\
\midrule
\multicolumn{1}{c|}{\multirow[c]{2}{*}{Semantics}}
& \multicolumn{1}{c|}{Cos.}                    & 0.0618 & 0.1000 & 0.0344 & 0.0446     & 0.0394 & 0.0584 & 0.0217 & 0.0266     & 0.0391 & 0.0520 & 0.0221 & 0.0256 \\
\multicolumn{1}{c|}{} & \multicolumn{1}{c|}{EASE}       & 0.0976 & 0.1536 & 0.0534 & 0.0683     & 0.0725 & 0.1044 & 0.0399 & 0.0483     & 0.0847 & 0.1198 & 0.0500 & 0.0598 \\
\midrule
\multicolumn{1}{c|}{\multirow[c]{4}{*}{Interaction}}
& \multicolumn{1}{c|}{EASE}                             & 0.1701 & 0.2448 & 0.0972 & 0.1172     & 0.0949 & 0.1260 & 0.0562 & 0.0645     & 0.1702 & 0.2241 & 0.1084 & 0.1236 \\
\multicolumn{1}{c|}{} & \multicolumn{1}{c|}{GF-CF}      & 0.1746 & 0.2470 & 0.0999 & 0.1195     & 0.0957 & 0.1307 & 0.0569 & 0.0663     & 0.1542 & 0.2132 & 0.0942 & 0.1108 \\
\multicolumn{1}{c|}{} & \multicolumn{1}{c|}{BSPM}       & 0.1760 & 0.2497 & 0.1017 & 0.1218     & 0.0956 & 0.1286 & 0.0578 & 0.0666     & 0.1596 & 0.2181 & 0.0996 & 0.1160 \\
\multicolumn{1}{c|}{} & \multicolumn{1}{c|}{SGFCF}      & 0.1855 & 0.2651 & 0.1072 & 0.1285     & 0.0993 & 0.1361 & 0.0587 & 0.0685     & 0.1691 & 0.2302 & 0.1055 & 0.1226 \\
\midrule
\multicolumn{1}{c|}{\multirow[c]{2}{*}{\makecell[c]{Interaction\\+ Multi-hot}}}
& \multicolumn{1}{c|}{CEASE}                            & 0.1730 & 0.2501 & 0.0987 & 0.1193     & 0.1065 & 0.1474 & 0.0624 & 0.0733     & 0.1714 & 0.2285 & 0.1070 & 0.1231 \\
\multicolumn{1}{c|}{} & \multicolumn{1}{c|}{Add-EASE}   & 0.1784 & 0.2565 & 0.0978 & 0.1186     & 0.1071 & 0.1462 & 0.0617 & 0.0722     & 0.1608 & 0.2284 & 0.0918 & 0.1109 \\
\midrule
\rowcolor[HTML]{FFF2CC} 
\multicolumn{1}{c|}{\multirow{1}{*}{\begin{tabular}[c]{@{}c@{}}Int. + Sem.\end{tabular}}} & \multicolumn{1}{c|}{L$^3$AE} & \textbf{0.1966*} & \textbf{0.2737*} & \textbf{0.1128*} & \textbf{0.1335*} & \textbf{0.1168*} & \textbf{0.1573*} & \textbf{0.0701*}  & \textbf{0.0810*} & \textbf{0.1818*}      & \textbf{0.2409*}      & \textbf{0.1151*}      & \textbf{0.1315*}    \\
\bottomrule
\end{tabular}
\end{adjustbox}
\end{table*}

%% file: Tables/Tab3_Fusion_Performance.tex
\begin{table}[t] \small
\caption{Performance over fusion methods on three datasets. LLM-CEASE and LLM-Add-EASE replace the tag-item matrix in CEASE and Add-EASE with L$^3$AE's semantic-item matrix.}
\centering
\label{tab:fusion_performance}
\renewcommand{\arraystretch}{0.9}
\begin{tabular}{c|c|cccc}
\toprule
Dataset & Model & R@10 & R@20 & N@10 & N@20 \\
\midrule
\multicolumn{1}{c|}{\multirow{3}{*}{Games}} & \multicolumn{1}{c|}{LLM-CEASE} & 0.1937 & 0.2687 & 0.1111 & 0.1313 \\
\multicolumn{1}{c|}{} & \multicolumn{1}{c|}{LLM-Add-EASE} & 0.1929 & 0.2712 & 0.1115 & 0.1327 \\
\multicolumn{1}{c|}{} & \multicolumn{1}{c|}{L$^3$AE} & \textbf{0.1966} & \textbf{0.2737} & \textbf{0.1128} & \textbf{0.1335} \\
\midrule
\multicolumn{1}{c|}{\multirow{3}{*}{Toys}} & \multicolumn{1}{c|}{LLM-CEASE} & 0.1144 & 0.1556 & 0.0681 & 0.0791 \\
\multicolumn{1}{c|}{} & \multicolumn{1}{c|}{LLM-Add-EASE} & 0.1136 & 0.1505 & 0.0685 & 0.0783 \\
\multicolumn{1}{c|}{} & \multicolumn{1}{c|}{L$^3$AE} & \textbf{0.1168} & \textbf{0.1573} & \textbf{0.0701} & \textbf{0.0810} \\
\midrule
\multicolumn{1}{c|}{\multirow{3}{*}{Books}} & \multicolumn{1}{c|}{LLM-CEASE} & 0.1800 & 0.2401 & 0.1135 & 0.1303 \\
\multicolumn{1}{c|}{} & \multicolumn{1}{c|}{LLM-Add-EASE} & 0.1802 & 0.2386 & 0.1140 & 0.1305 \\
\multicolumn{1}{c|}{} & \multicolumn{1}{c|}{L$^3$AE} & \textbf{0.1818} & \textbf{0.2409} & \textbf{0.1151} & \textbf{0.1315} \\
\bottomrule
\end{tabular}
\end{table}

%% file: Figures_tex/Figure3_Hyperparameter_Sensitivity.tex
\begin{figure}[t!]
\centering
\begin{tabular}{ccc}
\includegraphics[width=0.14\textwidth]{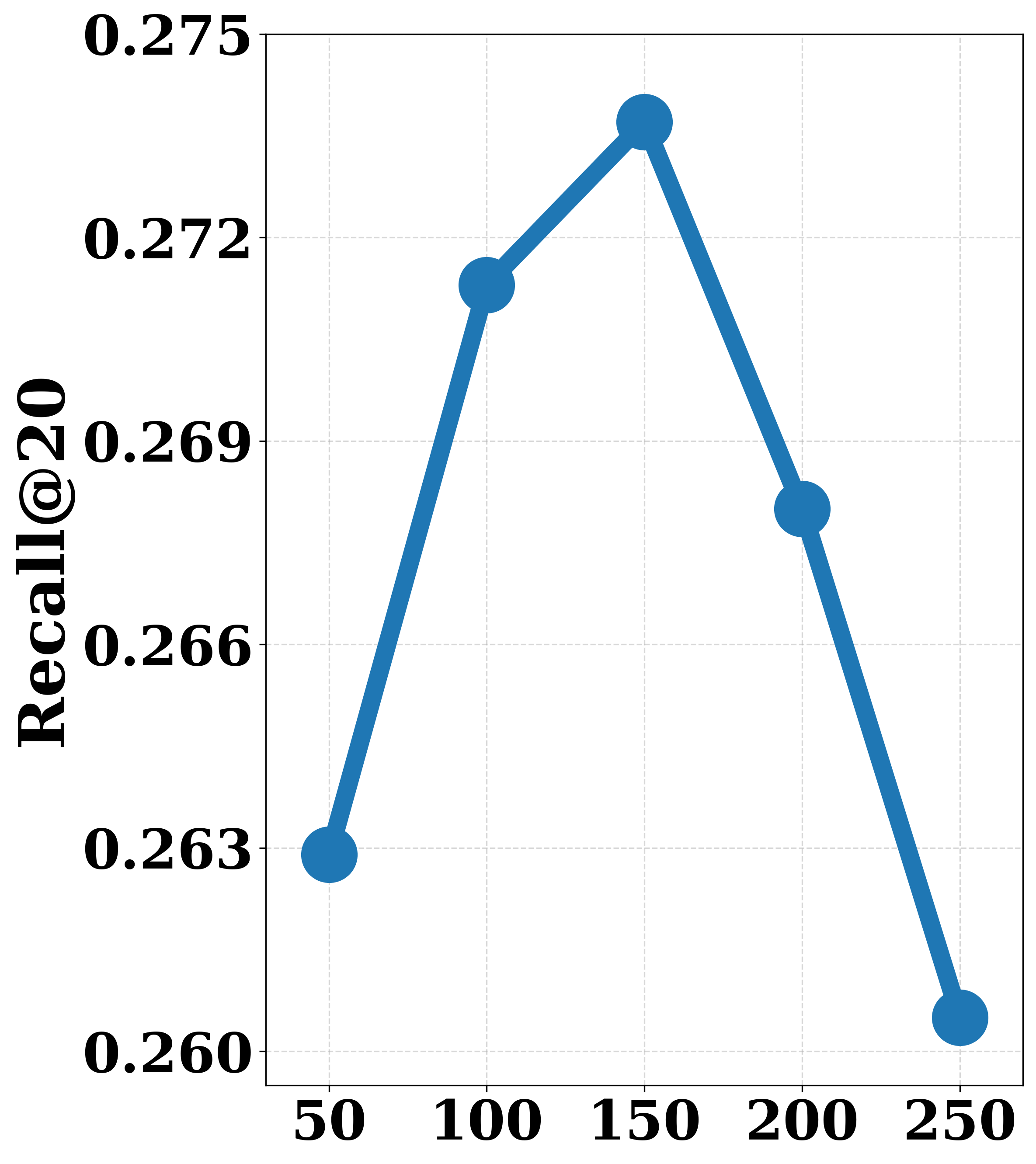} &
\includegraphics[width=0.14\textwidth]{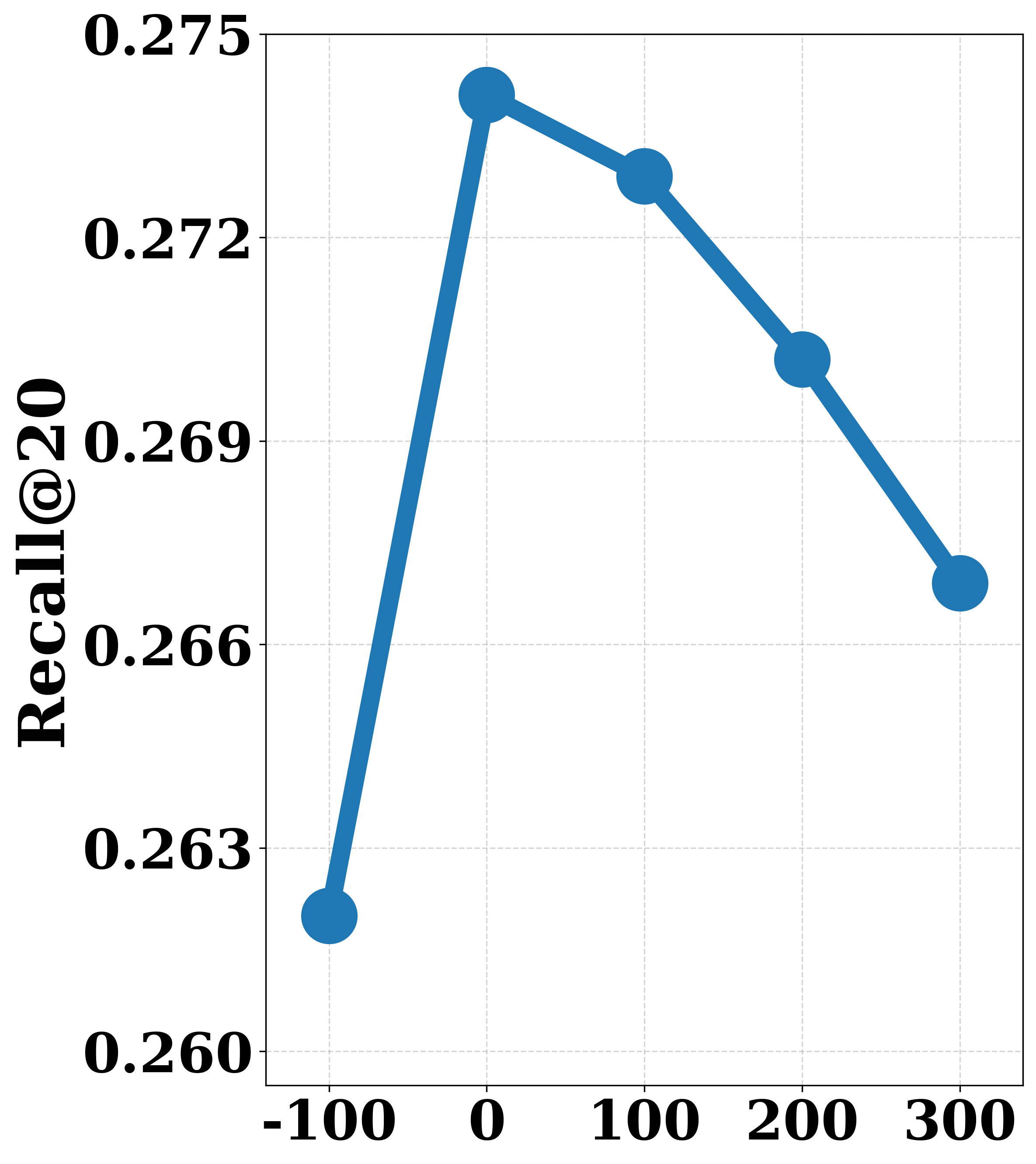} &
\includegraphics[width=0.14\textwidth]{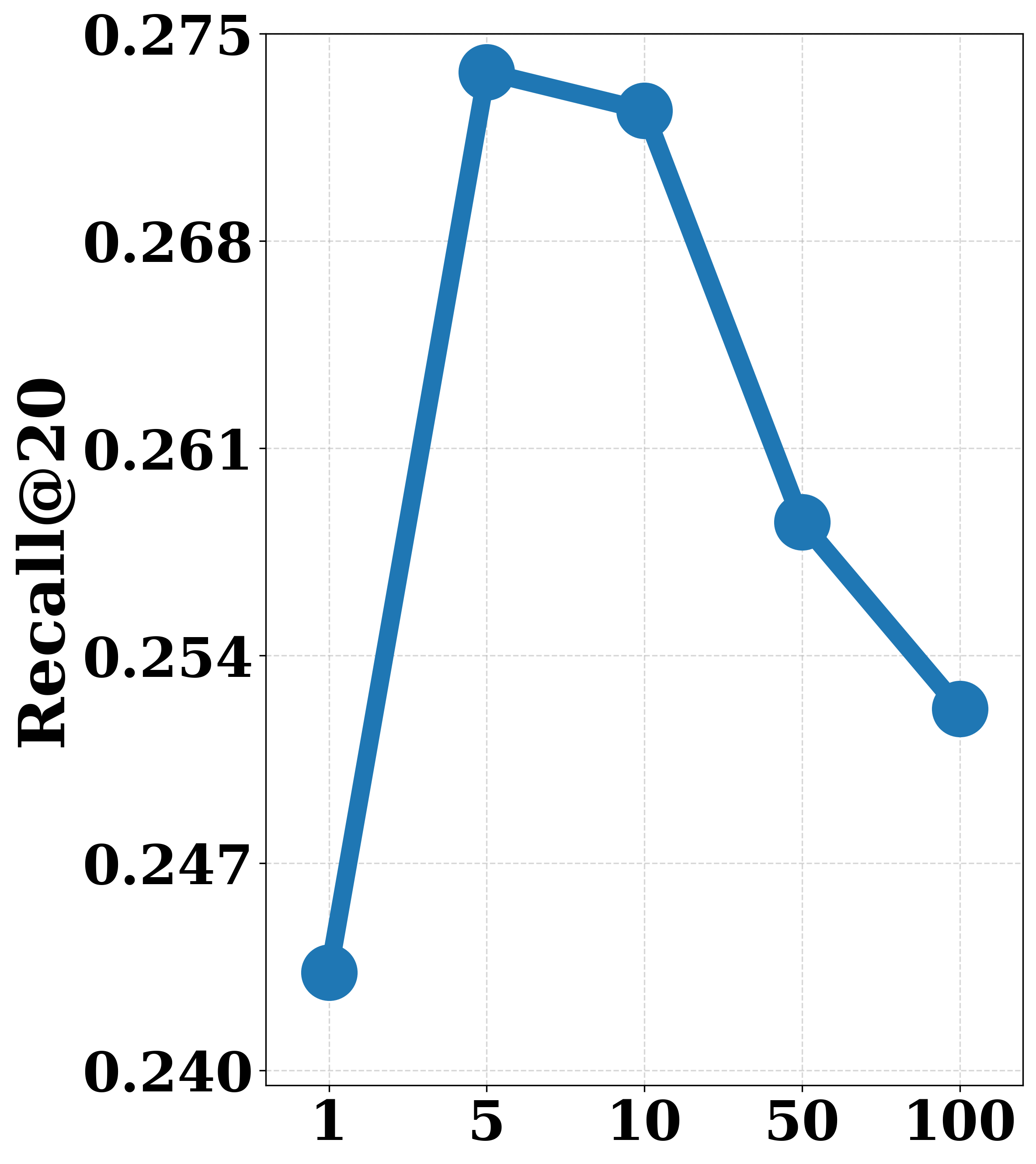} \\
(a) $\lambda_{KD}$ & (b) $\lambda_X$ & (c) $\lambda_F$ \\
\end{tabular}
\caption{Performance of L$^3$AE with varying L$_2$ regularization weights $\lambda_{KD}$, $\lambda_X$, and $\lambda_F$ on Games.}
\label{fig:hyper_sensitivity}
\end{figure}

%% file: Figures_tex/Figure4_Performance_over_Various_LLMs.tex
\begin{figure}[t!]
\centering
\begin{tabular}{c}
\includegraphics[width=0.35\textwidth]{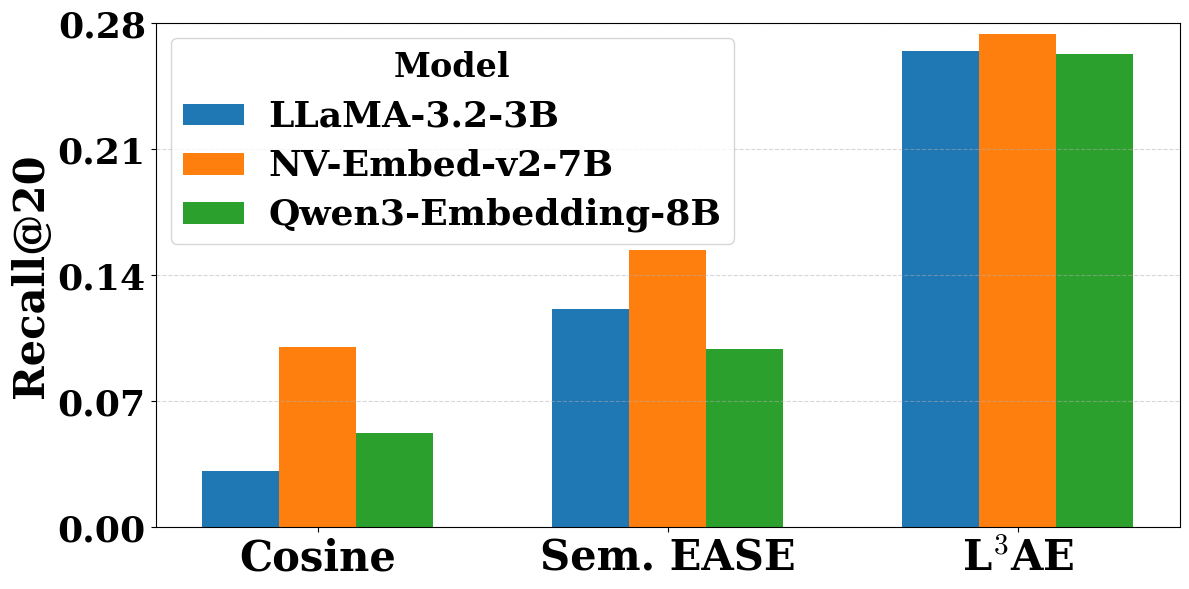} \\ 
(a) Games \\[3mm]
\includegraphics[width=0.35\textwidth]{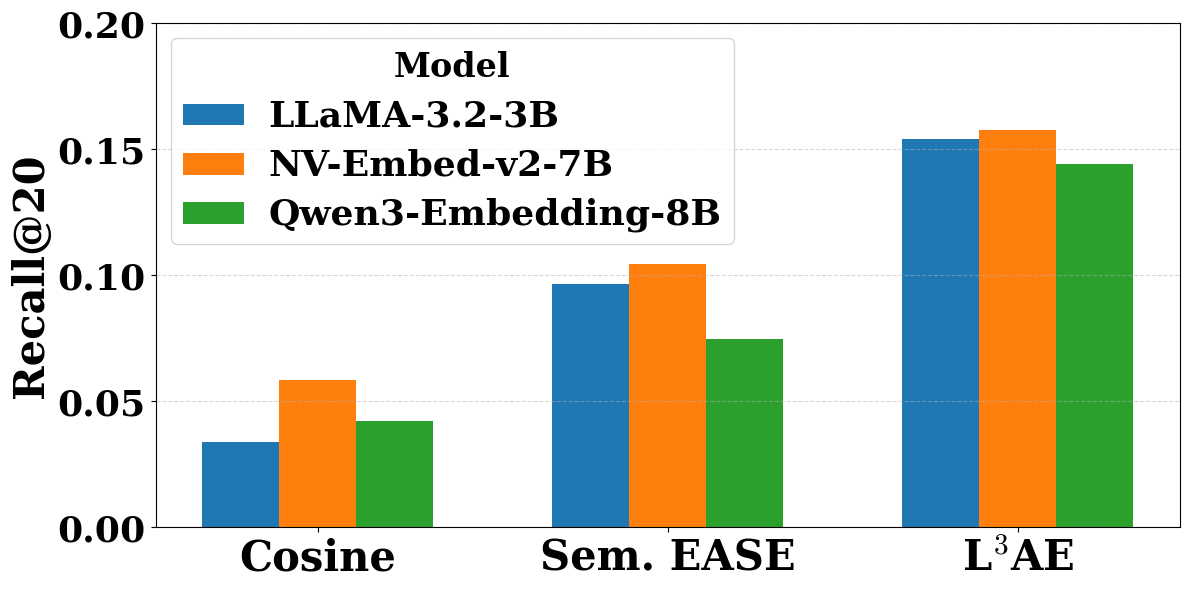} \\ 
(b) Toys \\[3mm]
\includegraphics[width=0.35\textwidth]{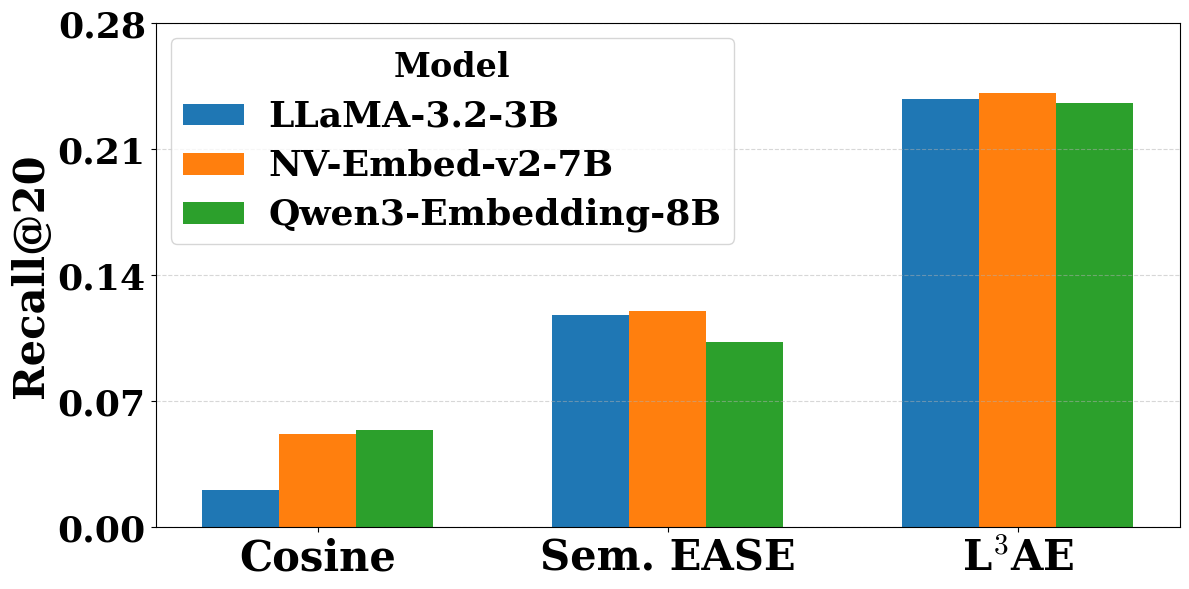} \\ 
(c) Books \\
\end{tabular}
\caption{Performance of cosine similarity, semantic-only EASE, and L$^3$AE with varying the LLM backbones (\ie, LLaMA-3.2-3B, NV-Embed-V2-7B, and Qwen3-Embedding-8B) on Games, Toys, and Books.}
\label{fig:performance_over_various_LLMs}
\end{figure}

%% file: Tables/Tab4_Overall_Performance_LLaMA.tex
\begin{table*}[t]
\caption{Performance comparison across three datasets with the LLaMA-3.2-3B backbone model. Bold indicates the best performance within each model category.}
\centering
\begin{adjustbox}{width=1\textwidth}
\label{tab:overall_llama}
\begin{tabular}{cccccc|cccc|cccc}
\toprule
\multicolumn{1}{c|}{\multirow{2}{*}{{\makecell[c]{Training \\ Features}}}} & \multicolumn{1}{c|}{\multirow{2}{*}{{Model}}}
& \multicolumn{4}{c|}{{Games}} & \multicolumn{4}{c|}{{Toys}} & \multicolumn{4}{c}{{Books}} \\
\multicolumn{1}{c|}{} & \multicolumn{1}{c|}{} & R@10 & R@20 & N@10 & N@20   & R@10 & R@20 & N@10 & N@20     & R@10 & R@20 & N@10 & N@20 \\
\midrule
\multicolumn{14}{c}{\textit{{Non-linear recommendation models}}} \\
\midrule
\multicolumn{1}{c|}{\multirow[c]{2}{*}{Interaction}}
& \multicolumn{1}{c|}{LightGCN}                         & 0.1453 & 0.2199 & 0.0799 & 0.0997     & 0.0520 & 0.0811 & 0.0281 & 0.0359     & 0.0973 & 0.1456 & 0.0566 & 0.0701 \\
\multicolumn{1}{c|}{} & \multicolumn{1}{c|}{SimGCL}     & 0.1510 & 0.2286 & 0.0831 & 0.1037     & 0.0611 & 0.0914 & 0.0338 & 0.0419     & 0.1122 & 0.1631 & 0.0661 & 0.0803 \\
\midrule
\multicolumn{1}{c|}{\multirow[c]{3}{*}{\makecell[c]{Interaction\\+ Semantics}}}
& \multicolumn{1}{c|}{RLMRec-Con} & \textbf{0.1606} & \textbf{0.2433} & \textbf{0.0898} & \textbf{0.1118}     & 0.0445 & 0.0701 & 0.0244 & 0.0311     & 0.1040 & 0.1518 & 0.0618 & 0.0752 \\
\multicolumn{1}{c|}{} & \multicolumn{1}{c|}{RLMRec-Gen} & 0.1524 & 0.2319 & 0.0848 & 0.1054     & 0.0637 & 0.0978 & 0.0347 & 0.0438     & 0.1124 & 0.1646 & \textbf{0.0665} & \textbf{0.0811} \\
\multicolumn{1}{c|}{} & \multicolumn{1}{c|}{AlphaRec}   & 0.1396 & 0.2108 & 0.0778 & 0.0966     & \textbf{0.0705} & \textbf{0.1042} & \textbf{0.0383} & \textbf{0.0472}     & \textbf{0.1144} & \textbf{0.1673} & 0.0659 & 0.0808 \\
\midrule
\multicolumn{14}{c}{\textit{{Linear recommendation models}}} \\
\midrule
\multicolumn{1}{c|}{\multirow[c]{2}{*}{Semantics}}
& \multicolumn{1}{c|}{Cos.}                             & 0.0211 & 0.0313 & 0.0108 & 0.0136     & 0.0227 & 0.0338 & 0.0125 & 0.0154     & 0.0152 & 0.0207 & 0.0093 & 0.0108 \\
\multicolumn{1}{c|}{} & \multicolumn{1}{c|}{EASE}       & 0.0774 & 0.1212 & 0.0407 & 0.0523     & 0.0656 & 0.0964 & 0.0372 & 0.0454     & 0.0778 & 0.1180 & 0.0437 & 0.0548 \\
\midrule
\multicolumn{1}{c|}{\multirow[c]{4}{*}{Interaction}}
& \multicolumn{1}{c|}{EASE}                             & 0.1701 & 0.2448 & 0.0972 & 0.1172     & 0.0949 & 0.1260 & 0.0562 & 0.0645     & 0.1702 & 0.2241 & 0.1084 & 0.1236 \\
\multicolumn{1}{c|}{} & \multicolumn{1}{c|}{GF-CF}      & 0.1746 & 0.2470 & 0.0999 & 0.1195     & 0.0957 & 0.1307 & 0.0569 & 0.0663     & 0.1542 & 0.2132 & 0.0942 & 0.1108 \\
\multicolumn{1}{c|}{} & \multicolumn{1}{c|}{BSPM}       & 0.1760 & 0.2497 & 0.1017 & 0.1218     & 0.0956 & 0.1286 & 0.0578 & 0.0666     & 0.1596 & 0.2181 & 0.0996 & 0.1160 \\
\multicolumn{1}{c|}{} & \multicolumn{1}{c|}{SGFCF}      & 0.1855 & \textbf{0.2651} & 0.1072 & 0.1285     & 0.0993 & 0.1361 & 0.0587 & 0.0685     & 0.1691 & 0.2302 & 0.1055 & 0.1226 \\
\midrule
\multicolumn{1}{c|}{\multirow[c]{2}{*}{\makecell[c]{Interaction\\+ Multi-hot}}}
& \multicolumn{1}{c|}{CEASE}                            & 0.1730 & 0.2501 & 0.0987 & 0.1193     & 0.1065 & 0.1474 & 0.0624 & 0.0733     & 0.1714 & 0.2285 & 0.1070 & 0.1231 \\
\multicolumn{1}{c|}{} & \multicolumn{1}{c|}{Add-EASE}   & 0.1784 & 0.2565 & 0.0978 & 0.1186     & 0.1071 & 0.1462 & 0.0617 & 0.0722     & 0.1608 & 0.2284 & 0.0918 & 0.1109 \\
\midrule
\rowcolor[HTML]{FFF2CC} 
\multicolumn{1}{c|}{\multirow{1}{*}{\begin{tabular}[c]{@{}c@{}}Int. + Sem.\end{tabular}}} & \multicolumn{1}{c|}{L$^3$AE}
& \textbf{0.1878} & 0.2641 & \textbf{0.1083} & \textbf{0.1288} & \textbf{0.1139} & \textbf{0.1540} & \textbf{0.0674}  & \textbf{0.0781} & \textbf{0.1797} & \textbf{0.2376} & \textbf{0.1137} & \textbf{0.1300}    \\
\bottomrule
\end{tabular}
\end{adjustbox}
\end{table*}

%% file: Tables/Tab5_Overall_Performance_Qwen3.tex
\begin{table*}[t]
\caption{Performance comparison across three datasets with the Qwen3-Embedding-8B backbone model. Bold indicates the best performance within each model category.}
\centering
\begin{adjustbox}{width=1\textwidth}
\label{tab:overall_qwen}
\begin{tabular}{cccccc|cccc|cccc}
\toprule
\multicolumn{1}{c|}{\multirow{2}{*}{{\makecell[c]{Training \\ Features}}}} & \multicolumn{1}{c|}{\multirow{2}{*}{{Model}}}
& \multicolumn{4}{c|}{{Games}} & \multicolumn{4}{c|}{{Toys}} & \multicolumn{4}{c}{{Books}} \\
\multicolumn{1}{c|}{} & \multicolumn{1}{c|}{} & R@10 & R@20 & N@10 & N@20   & R@10 & R@20 & N@10 & N@20     & R@10 & R@20 & N@10 & N@20 \\
\midrule
\multicolumn{14}{c}{\textit{{Non-linear recommendation models}}} \\
\midrule
\multicolumn{1}{c|}{\multirow[c]{2}{*}{Interaction}}
& \multicolumn{1}{c|}{LightGCN}                         & 0.1453 & 0.2199 & 0.0799 & 0.0997     & 0.0520 & 0.0811 & 0.0281 & 0.0359     & 0.0973 & 0.1456 & 0.0566 & 0.0701 \\
\multicolumn{1}{c|}{} & \multicolumn{1}{c|}{SimGCL}     & 0.1510 & 0.2286 & 0.0831 & 0.1037     & 0.0611 & 0.0914 & 0.0338 & 0.0419     & 0.1122 & 0.1631 & 0.0661 & 0.0803 \\
\midrule
\multicolumn{1}{c|}{\multirow[c]{3}{*}{\makecell[c]{Interaction\\+ Semantics}}}
& \multicolumn{1}{c|}{RLMRec-Con} & 0.1655 & 0.2406 & 0.0908 & 0.1108     & \textbf{0.0691} & \textbf{0.1052} & \textbf{0.0387} & \textbf{0.0483}     & 0.1127 & 0.1629 & 0.0663 & 0.0804 \\
\multicolumn{1}{c|}{} & \multicolumn{1}{c|}{RLMRec-Gen} & 0.1577 & 0.2356 & 0.0863 & 0.1072     & 0.0674 & 0.1014 & 0.0376 & 0.0467     & 0.1148 & 0.1657 & 0.0681 & 0.0824 \\
\multicolumn{1}{c|}{} & \multicolumn{1}{c|}{AlphaRec}   & \textbf{0.1707} & \textbf{0.2546} & \textbf{0.0930} & \textbf{0.1154}     & 0.0688 & 0.1029 & 0.0379 & 0.0470     & \textbf{0.1239} & \textbf{0.1729} & \textbf{0.0730} & \textbf{0.0867} \\
\midrule
\multicolumn{14}{c}{\textit{{Linear recommendation models}}} \\
\midrule
\multicolumn{1}{c|}{\multirow[c]{2}{*}{Semantics}}
& \multicolumn{1}{c|}{Cos.}                             & 0.0343 & 0.0523 & 0.0201 & 0.0249     & 0.0289 & 0.0423 & 0.0156 & 0.0192     & 0.0392 & 0.0539 & 0.0229 & 0.0268 \\
\multicolumn{1}{c|}{} & \multicolumn{1}{c|}{EASE}       & 0.0620 & 0.0988 & 0.0330 & 0.0429     & 0.0493 & 0.0747 & 0.0280 & 0.0347     & 0.0692 & 0.1027 & 0.0401 & 0.0493 \\
\midrule
\multicolumn{1}{c|}{\multirow[c]{4}{*}{Interaction}}
& \multicolumn{1}{c|}{EASE}                             & 0.1701 & 0.2448 & 0.0972 & 0.1172     & 0.0949 & 0.1260 & 0.0562 & 0.0645     & 0.1702 & 0.2241 & 0.1084 & 0.1236 \\
\multicolumn{1}{c|}{} & \multicolumn{1}{c|}{GF-CF}      & 0.1746 & 0.2470 & 0.0999 & 0.1195     & 0.0957 & 0.1307 & 0.0569 & 0.0663     & 0.1542 & 0.2132 & 0.0942 & 0.1108 \\
\multicolumn{1}{c|}{} & \multicolumn{1}{c|}{BSPM}       & 0.1760 & 0.2497 & 0.1017 & 0.1218     & 0.0956 & 0.1286 & 0.0578 & 0.0666     & 0.1596 & 0.2181 & 0.0996 & 0.1160 \\
\multicolumn{1}{c|}{} & \multicolumn{1}{c|}{SGFCF}      & \textbf{0.1855} & \textbf{0.2651} & \textbf{0.1072} & \textbf{0.1285}     & 0.0993 & 0.1361 & 0.0587 & 0.0685     & 0.1691 & 0.2302 & 0.1055 & 0.1226 \\
\midrule
\multicolumn{1}{c|}{\multirow[c]{2}{*}{\makecell[c]{Interaction\\+ Multi-hot}}}
& \multicolumn{1}{c|}{CEASE}                            & 0.1730 & 0.2501 & 0.0987 & 0.1193     & 0.1065 & \textbf{0.1474} & 0.0624 & \textbf{0.0733}     & 0.1714 & 0.2285 & 0.1070 & 0.1231 \\
\multicolumn{1}{c|}{} & \multicolumn{1}{c|}{Add-EASE}   & 0.1784 & 0.2565 & 0.0978 & 0.1186     & \textbf{0.1071} & 0.1462 & 0.0617 & 0.0722     & 0.1608 & 0.2284 & 0.0918 & 0.1109 \\
\midrule
\rowcolor[HTML]{FFF2CC} 
\multicolumn{1}{c|}{\multirow{1}{*}{\begin{tabular}[c]{@{}c@{}}Int. + Sem.\end{tabular}}} & \multicolumn{1}{c|}{L$^3$AE}
& 0.1822 & 0.2625 & 0.1042 & 0.1257 & 0.1060 & 0.1441 & \textbf{0.0627}  & 0.0729 & \textbf{0.1790} & \textbf{0.2356} & \textbf{0.1139} & \textbf{0.1299}    \\
\bottomrule
\end{tabular}
\end{adjustbox}
\end{table*}

%% file: sec-conclusion.tex
\section{Conclusion}\label{sec:conclusion}

This paper explored the first integration of LLMs into LAEs for CF. We demonstrated that semantic embeddings generated by LLMs completely supplant traditional multi-hot encoding schemes. To effectively integrate heterogeneous knowledge from both textual semantics and interaction data, we propose L$^3$AE with a two-phase optimization, guaranteeing a globally optimal closed-form solution. L$^3$AE outperformed state-of-the-art LLM-enhanced methods on three datasets, establishing that LLM-enhanced linear architectures can be an effective alternative to complex neural CF models. The source code is available at \url{https://github.com/jaewan7599/L3AE_CIKM2025}.